# Circularly polarized electroluminescence from silicon nanostructures heavily doped with boron

N.T. Bagraev, L.E. Klyachkin, R.V. Kuzmin[a)], A.M. Malyarenko

*Ioffe Physical-Technical Institute of the Russian Academy of Sciences, St.Petersburg, 194021, Russia*

[a)]Electronic mail: roman.kuzmin@mail.ioffe.ru

**Abstract**

The circularly polarized electroluminescence (CPEL) from silicon nanostructures which are the p-type ultra-narrow silicon quantum well (Si-QW) confined by δ-barriers heavily doped with boron, $5 \times 10^{21}$ cm$^{-3}$, is under study as a function of temperature and excitation levels. The CPEL dependences on the forward current and temperature show the circularly polarized light emission which appears to be caused by the exciton recombination through the negative-U dipole boron centers at the Si-QW δ-barriers interface.

Possibility to create a fully silicon-compatible optoelectronics attracts a lot of attention to the problem of silicon light emission source.[1] Numerous approaches, such as porous Si, Si nanocrystals, erbium doped Si, plastically deformed monocrystalline silicon were prepared to obtain an effective luminescence from silicon-based material.[2-6] Last decade, many observations of the high intensive near band-edge electroluminescence (EL) from the silicon p-n junctions have been documented.[7-10] This quite surprising result for indirect band gap semiconductors was commonly accounted for by carrier confinement in the dislocation lops introduced by the boron ion implantation process. However the high intensity of the near band-edge electroluminescence was also observed from the p-n junctions prepared by boron diffusion, which seem to avoid the formation of the dislocation loops.[11] Furthermore this emission was found to be not related directly to the presence of the dislocations which may affect only as a getter of nonradiative recombination centers.[12]

Nevertheless there is a common feature of the most works used either ion implantation or diffusion of boron. This feature is rather high level of boron doping, when its ordinary concentration is near the solubility limit. Under these conditions the formation of the locally enhanced boron doping spikes appears to take place.[13,14] Such spikes were shown to confine excitons effectively and thus provide the intensive near band-edge electroluminescence by releasing free electron-hole pairs at elevated temperatures.[15] The EL intensity is dependent on the concentration of boron reaching a maximum at $N(B) \sim 4\times10^{20}$ cm$^{-3}$.[15] The formation of the boron clusters which is unavoidable process within frameworks of the standard impurity doping techniques is able to prevent further EL intensity enhancement.

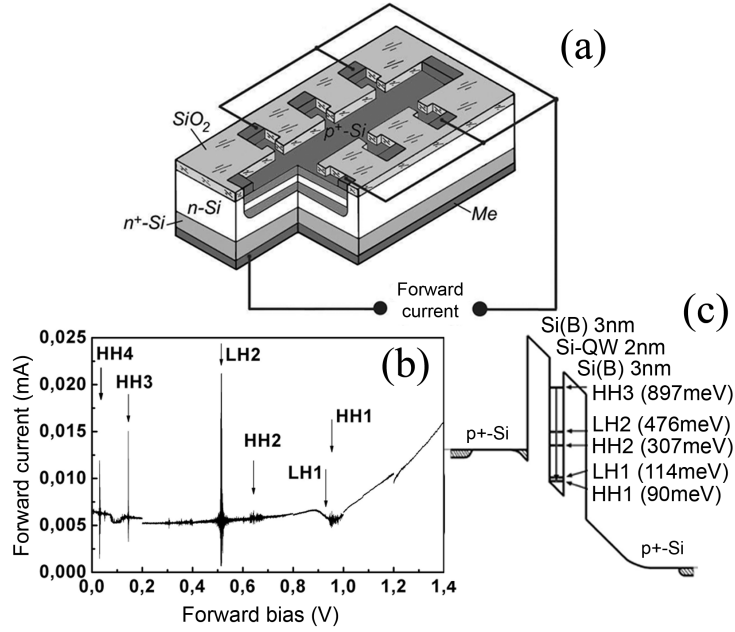

Fig. 1. (a) A schematic diagram of the device that demonstrates a perspective view of the p-type Si-QW confined by the δ-barriers heavily doped with boron on the n-type Si (100) surface. (b) The forward current-voltage characteristic of the p+-n junction that defines the energies of the two-dimensional subbands of 2D holes in the p-type Si-QW confined by δ-barriers (c).

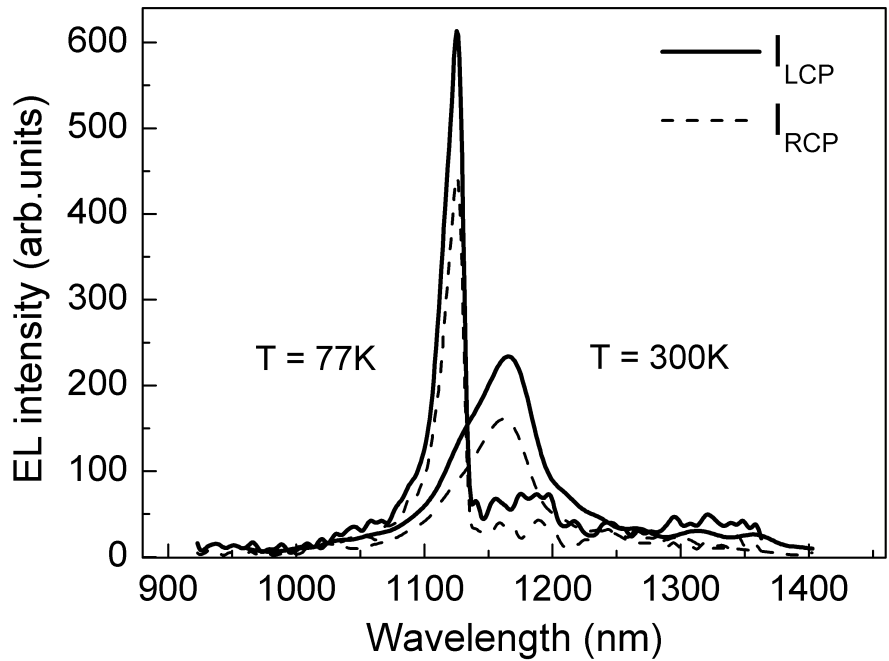

Fig. 2. Left- ($I_{LCP}$) and right-circularly polarized ($I_{RCP}$) EL spectra components of the p-type silicon nanostructure on the n-type Si (100) surface detected at the forward current of 20 mA for 77 K and 300 K.

However this limit appeared to be avoided if short-time vacancy enhanced diffusion of boron is used after preliminary oxidation of the n-type Si (100) surface. Under fine injection of vacancies during diffusion process the ultra-narrow Si-QW, 2 nm, has been established to be self-assembled between the δ-barriers heavily doped with boron in the concentration of $5\times10^{21}$ cm$^{-3}$, which is controlled by the secondary-ion mass spectroscopy (SIMS) method. Here we present the first findings of the near band-edge electroluminescence from these p-type silicon nanostructures on the n-type Si (100) surface. The electroluminescence appears to reveal rather high intensity even at room temperature and moreover the high degree of circular polarization. The CPEL mechanism seems to be started from the recombination of the excitons bounded by the negative-U dipole boron centers at the Si-QW - δ-barriers interface.

We used a 350 μm thick n-Si (100) wafer of 20 Ω×cm$^{-1}$ resistivity. The short-time diffusion of boron was performed at the temperature of 900°C after preliminary oxidation in the presence of dry oxygen and chlorine compounds as well as subsequent photolithography and etching. The depth of the ultra-shallow boron diffusion profile determined by SIMS method was 8 nm. For further electroluminescence measurements the gold contacts were prepared on the front surface of the sample using standard photolithography technique, whereas the back side of the substrate was covered by aluminum. The device was designed within frameworks of the Hall geometry with the doping area of size 4.7×0.1 mm which is provided by eight 200×200 μm contacts as shown in Fig. 1a. The forward current-voltage characteristic of the p$^+$-n junction prepared is shown in Fig. 1b which reveals the system of energy levels inside the p-type Si-QW (Fig. 1c). The samples obtained were set up in a cryostat for the EL measurements in the temperature range from 4.2 to 300 K. The electroluminescence was detected with the MDR-23 monochromator using an InGaAs photodiode and a conventional lock-in technique.

The EL spectra detected reveal the left- and right-hand circular polarization with high degree up to 20% (see Fig. 2). The main emission line in the low-temperature EL spectrum is observed at 1126 nm with the full-width at half-maximum (FWHM) of 18 nm and the phonon replica at 1194 nm. Under elevated temperature, this line saves the circular polarization decreasing slightly in intensity and increasing in FWHM to 70 nm as well as shifting to 1163 nm following by the band gap temperature dependence. The intensity of the main emission line was calibrated with a gauged source in the same spectral range. The spatial distribution of the electroluminescence was taken into account. The EL power obtained is shown in Fig. 3 as a function of the forward current, which exhibits the linear character at temperatures of 77 K and 300 K except low values. It should be noted that the external quantum efficiency is difficult to be estimated due to the presence of many emission bands in very wide spectral range from visible to far infrared regions.[16,17] So it is hard to know which part of the input power participates in the emission under the study.

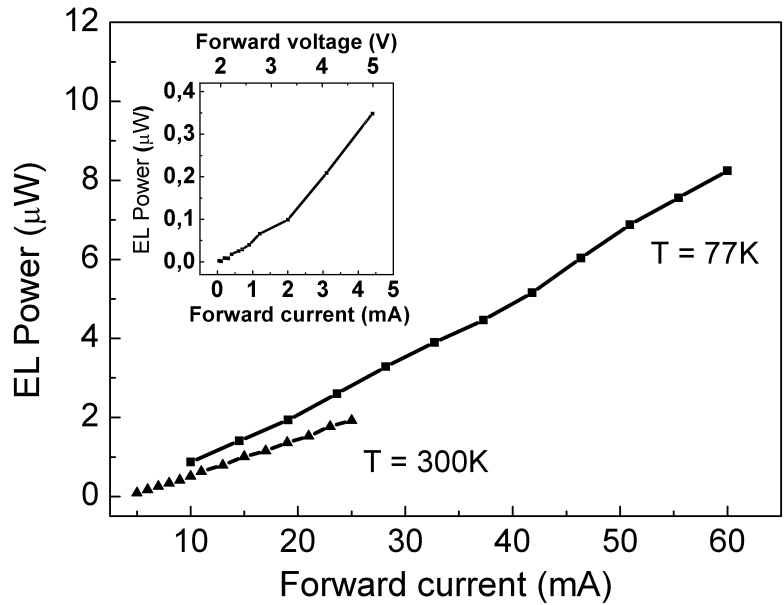

Fig. 3. The electroluminescence power as a function of forward current at various temperatures: 77 K (squares), 300 K (triangles). Inset: low current part of the function.

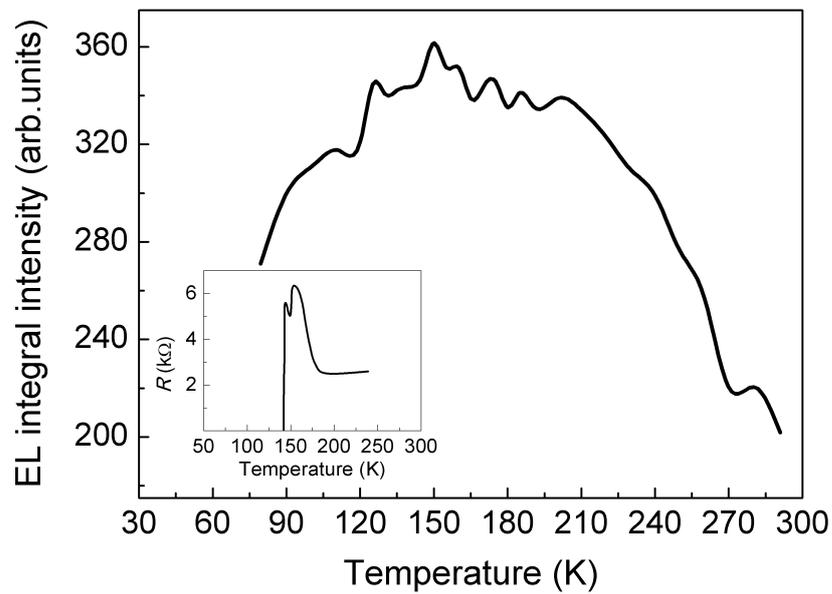

Fig. 4. The temperature dependence of the integral electroluminescence intensity of the main emission line revealing many features near 150 K which correlate with the value of the critical temperature of the superconductor transition, $T_c \sim 145$ K, estimated from electrical resistivity (see inset), thermo-emf and magnetic susceptibility measurements.

To clarify the CPEL mechanism, the structure of the boron ultra-shallow diffusion profiles should be defined. The cyclotron resonance (CR) and ESR measurements as well as the scanning tunneling microscopy (STM) technique were used. The CR measurements were performed at 3.8 K with a standard Brucker-Physik AG ESR spectrometer at X-band (9.1-9.5 GHz). The CR quenching and the angular dependence for which a characteristic 180° symmetry was observed revealed the confining potential inside $p^+$-diffusion profile and its different arrangement in longitudinal and lateral Si-QW s, which are formed naturally between the δ-barriers heavily doped with boron.[18,19] These Si-QW s were shown to contain the high mobility 2D hole gas which is characterized by long momentum relaxation time of heavy and light holes at 3.8 K.[18-20] The STM technique revealed the formation of the self-interstitials microdefects inside the δ-barriers confining Si-QW. The size of the smallest microdefect obtained is 2 nm, which is consistent with the parameters expected from the tetrahedral model of the $Si_{60}$ cluster.[21] The STM images of larger scales demonstrated that the ratio between the dimensions of the microdefects is supported to be equal to 3.3 thereby defining the self-assembly of the self-interstitials microdefects as the self-organization of the fractal type. Thus the δ-barriers heavily doped with boron, $5 \times 10^{21}$ cm$^{-3}$, represent alternating arrays of the smallest undoped microdefects and doped dots with dimensions restricted to 2 nm. The value of the boron concentration determined by the SIMS method seems to indicate that each doped dot located between undoped microdefects contains two impurity atoms of boron. To study of the boron centers packed up in these dots the ESR technique was applied. The same Brucker-Physik AG ESR spectrometer at X-band (9.1-9.5 GHz) was used with the rotation of the magnetic field in the {110}-plane perpendicular to a {100}-interface. The angular dependences of the ESR spectra showed that the doped microdefects appear to consist of the trigonal dipole boron centers, $B^+ - B^-$, which are caused by the elastic reconstruction along the [111] crystallographic axis of the shallow acceptors as a result of the negative-U reaction: $2B^0 \rightarrow B^+ + B^-$.[22] The trigonal ESR spectrum seems to be evidence of the dynamic magnetic moment that is induced by the exchange interaction between the small hole bipolarons which are formed by this negative-U reconstruction of the boron acceptors.[18,22] Thus the dipole boron centres are a basis of nanostructured δ - barriers confining the Si-QW. Furthermore, it was found by the electrical resistivity, thermo-emf and magnetic susceptibility measurements that the δ-barriers heavily doped with boron, $N(B) = 5 \times 10^{21}$ cm$^{-3}$, reveal a phase transition at critical temperature of $T_c \sim 145$

K which is accompanied with appearing of the superconductor properties caused by transferring of the small hole bipolarons through these negative-U dipole boron centers.[17]

It seems to be likely that the microdefects containing negative-U dipole centers of boron should to affect to the electroluminescence emitted from the silicon nanostructure prepared. As a proof of this assumption the temperature dependence of the electroluminescence intensity, as can be seen in Fig. 4, increases with the temperature until reaches a maximum in a vicinity of 150 K. This EL temperature dependence seems to be result from the exciton confinement caused by the chains of the negative-U dipole boron centers which collapse at elevated temperatures. The temperature of the EL intensity quenching is in good agreement with the value of the critical temperature of the superconductor transition, $T_c \sim 145$ K, estimated from electrical resistivity (see inset in Fig. 4), thermo-emf and magnetic susceptibility measurements. The fact of the circular polarization of the electroluminescence can also be referred to the negative-U reconstruction of the shallow boron acceptors, because the high temperature superconductor properties for the δ-barriers was found to be a result of coherent state of the small hole bipolarons on the dipole centers of boron.[17,22] In that case the high degree CPEL may be a result of breakdown of the coherent state of small hole bipolarons at these negative-U centers. Besides, Fig. 2 shows that the high degree CPEL is retained up to room temperature, which appears to be due to the conservation of the spin-lattice relaxation time as a function of temperature in ultra-narrow quantum wells, wires etc.[23]

The strong highly polarized electroluminescence from the p-type ultra-narrow Si-QW confined by the δ-barriers heavily doped with boron, $5 \times 10^{21}$ cm$^{-3}$, on the n-type Si (100) surface has been demonstrated. The EL studies on different temperatures and excitation levels have shown that the mechanism of the efficient circularly polarized electroluminescence seems to be started from the recombination of the excitons confined by the chains of the negative-U dipole boron centers.

The work was supported by the programme of fundamental studies of the Presidium of the Russian Academy of Sciences "Quantum Physics of Condensed Matter" (grant 9.12); programme of the Swiss National Science Foundation (grant IZ73Z0_127945/1); the Federal Targeted Programme on Research and Development in Priority Areas for the Russian Science and Technology Complex in 2007–2012 (contract no. 02.514.11.4074), the SEVENTH FRAMEWORK PROGRAMME Marie Curie Actions PIRSES-GA-2009-246784 project SPINMET.